\newcommand{\PaperTitle}{Learning Cellular Network Connection Quality with Conformal}
\newcommand{\PaperNumber}{XXX}

\documentclass[10pt,sigconf,letterpaper,nonacm]{acmart}
\author{Hanyang Jiang, Elizabeth Belding, Ellen Zegure, Yao Xie}
\usepackage[ruled,vlined]{algorithm2e}
\usepackage{bm}
\begin{document}

\title{\PaperTitle}

\begin{abstract}
In this paper, we address the problem of uncertainty quantification for cellular network speed. It is a well-known fact that the actual internet speed experienced by a mobile phone can fluctuate significantly, even when remaining in a single location. This high degree of variability underscores that mere point estimation of network speed is insufficient. Rather, it is advantageous to establish a prediction interval that can encompass the expected range of speed variations. In order to build an accurate network estimation map, numerous mobile data need to be collected at different locations. Currently, public datasets rely on users to upload data through apps. Although massive data has been collected, the datasets suffer from significant noise due to the nature of cellular networks and various other factors. Additionally, the uneven distribution of population density affects the spatial consistency of data collection, leading to substantial uncertainty in the network quality maps derived from this data. We focus our analysis on large-scale internet-quality datasets provided by Ookla to construct an estimated map of connection quality. To improve the reliability of this map, we introduce a novel conformal prediction technique to build an uncertainty map. We identify regions with heightened uncertainty to prioritize targeted, manual data collection. In addition, the uncertainty map quantifies how reliable the prediction is in different areas. Our method also leads to a sampling strategy that guides researchers to selectively gather high-quality data that best complement the current dataset to improve the overall accuracy of the prediction model.
\end{abstract}
\maketitle

\section{Introduction}
Since the advent of cellular-network-based wireless internet over three decades ago, mobile internet has become a cornerstone of daily life, facilitating a wide range of activities in sectors such as communication, entertainment, commerce, and education. While urban centers typically enjoy strong mobile internet connectivity due to well-developed infrastructure, rural areas frequently suffer from inadequate connectivity \cite{vogels2021some}. This discrepancy primarily stems from an insufficient number of cell towers and infrastructural challenges, which are influenced by factors like geographic spread, lower population density, and diminished financial incentives for investing in these areas. Accurately pinpointing these locations can significantly improve network planning \cite{niu2021greening}.

Identifying regions that are poorly served or entirely unserved is crucial for telecommunication operators and policymakers \cite{mangla2021tale}. By focusing on these areas, service providers can deliberately improve infrastructure to enhance both coverage and the quality of service. On a larger scale, detailed connectivity maps can guide policy decisions, promoting digital inclusion and helping to close the gap between urban and rural communities. Enhancing mobile internet access in rural areas also offers considerable socio-economic benefits, fostering regional development and inclusivity. Therefore, addressing the challenges of mobile internet access in under-connected regions is essential for fostering digital equity and propelling societal advancement. In the United States, the Broadband Equity, Access, and Deployment (BEAD) program is set to direct significant funding towards expansion in the coming years, influenced partly by data on spatial quality~\cite{BEAD}.

However, current mobile signal estimation work \cite{zhang2020cellular} often utilizes propagation models to assess connection quality. The cell tower locations are not shared by Internet service providers, and the public datasets are contributed mostly by volunteers. This makes the dataset noisy, inaccurate, and not up-to-date. Furthermore, the propagation models are limited in accuracy. In reality, the connection quality can be quite different either because of incomplete data or other contributing factors.

Another popular approach to estimating quality is to use data-motivated methods \cite{jiang2023mobile} like kernel regression or Kriging. These methods are able to reach higher accuracy but still suffer from noisy and incomplete datasets. The uneven distribution of the population results in insufficient data in rural and undeveloped areas, which are targets for investments. If we can identify those areas that require more data, we could manually collect high-quality data in those places and improve predictions.

In this work, we propose a novel method called Ensemble Spatial Conformal Prediction (ESCP) to identify those areas where the prediction models have higher uncertainty. The reason for a higher uncertainty in an area can be an inherent higher variation in a cellular network, lack of data, failure of the prediction model, and so on. The uncertainty quantification can identify these areas for further improvement. Our work is based on an in-depth analysis using one of the largest open datasets, namely the Ookla dataset \cite{ookla2022}, which captures mobile internet connectivity measurements worldwide. Users conduct speed tests on their internet connections using Ookla's tools. When a user initiates a speed test, data such as the user's IP address, the Internet Service Provider's (ISP) identity, and the connection's speed (both download and upload measured connection bandwidth) are logged. However, like many real-world datasets, it presents certain challenges. It is characterized by a relatively high noise and a conspicuous absence of measurements in specific areas.

Conformal prediction (CP) is a widely recognized distribution-free method used to quantify uncertainty in contemporary machine learning contexts \cite{vovkinductive}. This technique enhances conventional predictive algorithms by not only generating point estimates but also providing uncertainty intervals. These intervals are designed to encapsulate the unobserved ground truth with a high probability specified by the user. Consequently, CP has been effectively employed across various domains, including anomaly detection \cite{xu2021ECAD}, classification \cite{MJ_classification,xuERAPS2022}, and regression \cite{jackknife+}, among others.

Fundamentally, CP operates as a flexible wrapper around any predictive model, typically described as a black-box function $f$. It integrates three key elements: the model $f$, an input feature $X$, and a potential outcome $Y$. The core of CP involves computing a "non-conformity" score, which assesses the degree to which the potential outcome deviates from typical model predictions. This score is crucial for determining how well the potential outcome aligns with established patterns recognized by the model. Through this approach, CP provides a robust framework for uncertainty quantification in predictive modeling, enhancing the reliability and interpretability of machine learning applications.

For the prediction model $f$, we design a self-tuning bandwidth kernel regression method. The method is designed to deal with the spatial imbalance of data. In contrast to conventional models, the approach is designed to adaptively determine the kernel regression bandwidth for a location.

In this study, we aim to create a prediction region map of mobile internet quality that covers an entire state, not just urban areas seen in prior studies. Some works, including \cite{adarsh2021coverage}, realize the importance of this field, but they focus more on the data collection and analysis part instead of developing methods. The lack of data from rural areas makes this a tough challenge, setting our work apart from existing approaches. 

The rest of the paper is organized as follows. We first give a detailed description of the datasets used in the paper. After that, we introduce the methods used in this paper. Finally, we present the numerical experiments on the two different datasets to show the performance of our proposed method.

\section{Data Description}
The dataset utilized in our study is sourced from the open datasets provided by Ookla. The data gathered by Ookla from 2019 to 2022 encapsulates the performance metrics of mobile internet connections for a multitude of users worldwide. Key variables in this dataset include geographical coordinates (longitude and latitude), mean download speed (MB/s), mean upload speed (MB/s), count of tests conducted in each area (aggregated for user privacy into  $600m^2$ grid blocks), the number of distinct devices utilized for testing, and a comprehensive score assessing the connection speed. The public Ookla datasets do not provide information on the cellular provider, hence our estimation is of coverage quality aggregated over all providers present in the measurements. The availability of data labeled by provider would be of great interest to the research community.

\begin{figure}[t]
    \centering
    \includegraphics[scale=0.315]{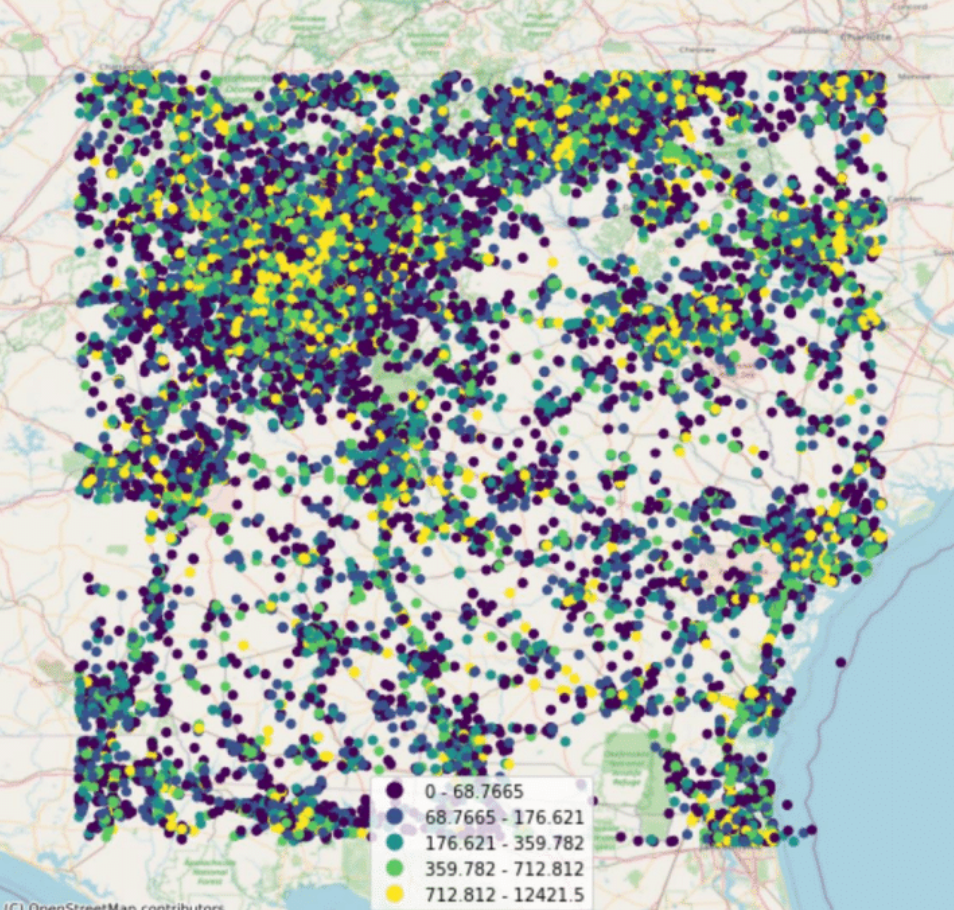}
    \includegraphics[scale=0.385]{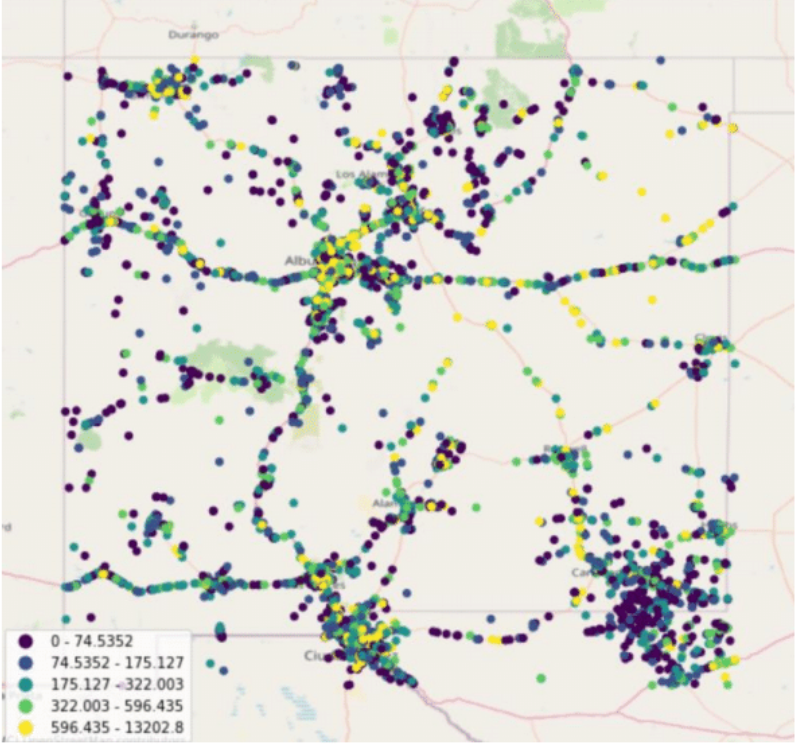}
    \caption{Mobile connection score of Georgia and New Mexico, where the data in Georgia is very dense, and in New Mexico is sparse.}
    \label{s1}
\end{figure}

\begin{figure}[t]
    \centering
    \includegraphics[scale=0.225]{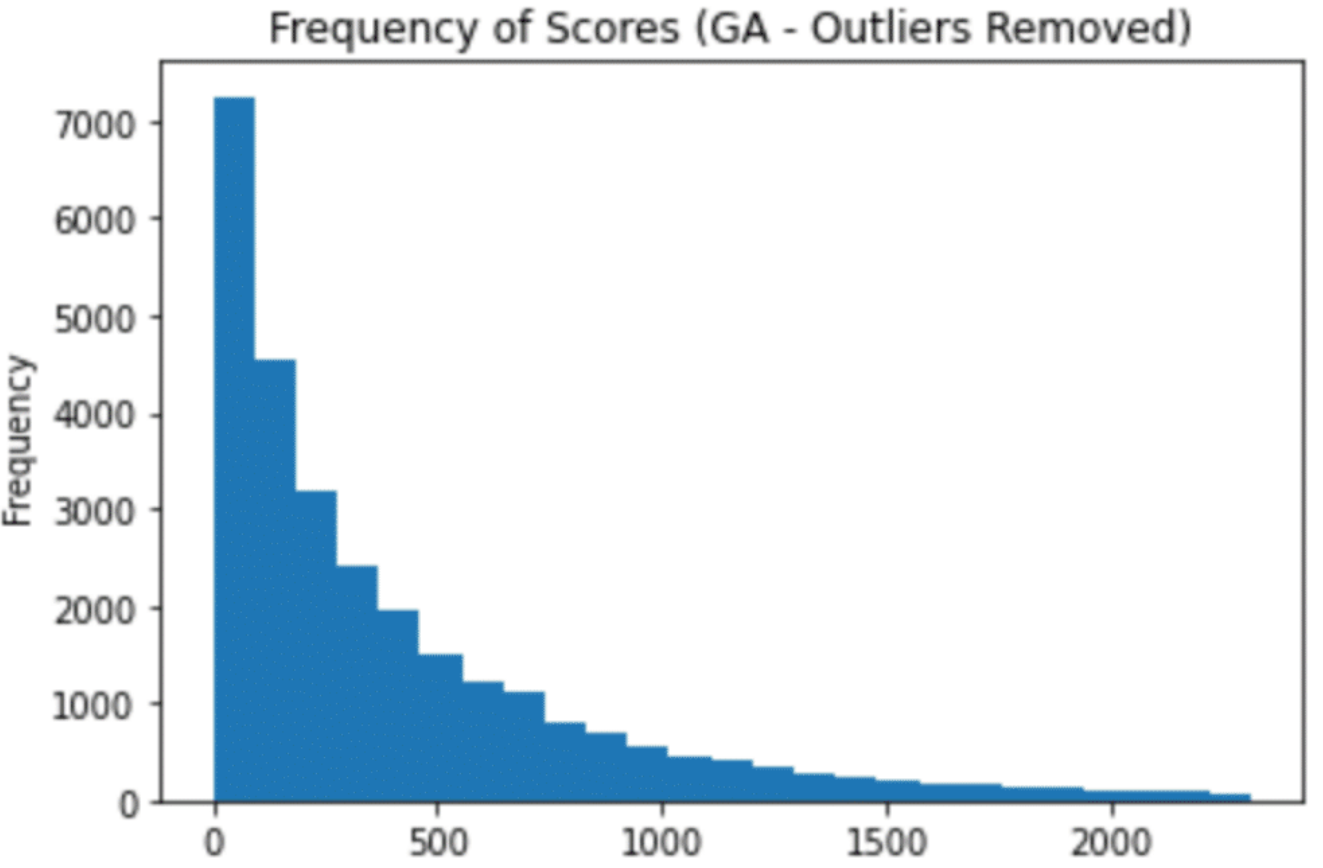}
    \includegraphics[scale=0.255]{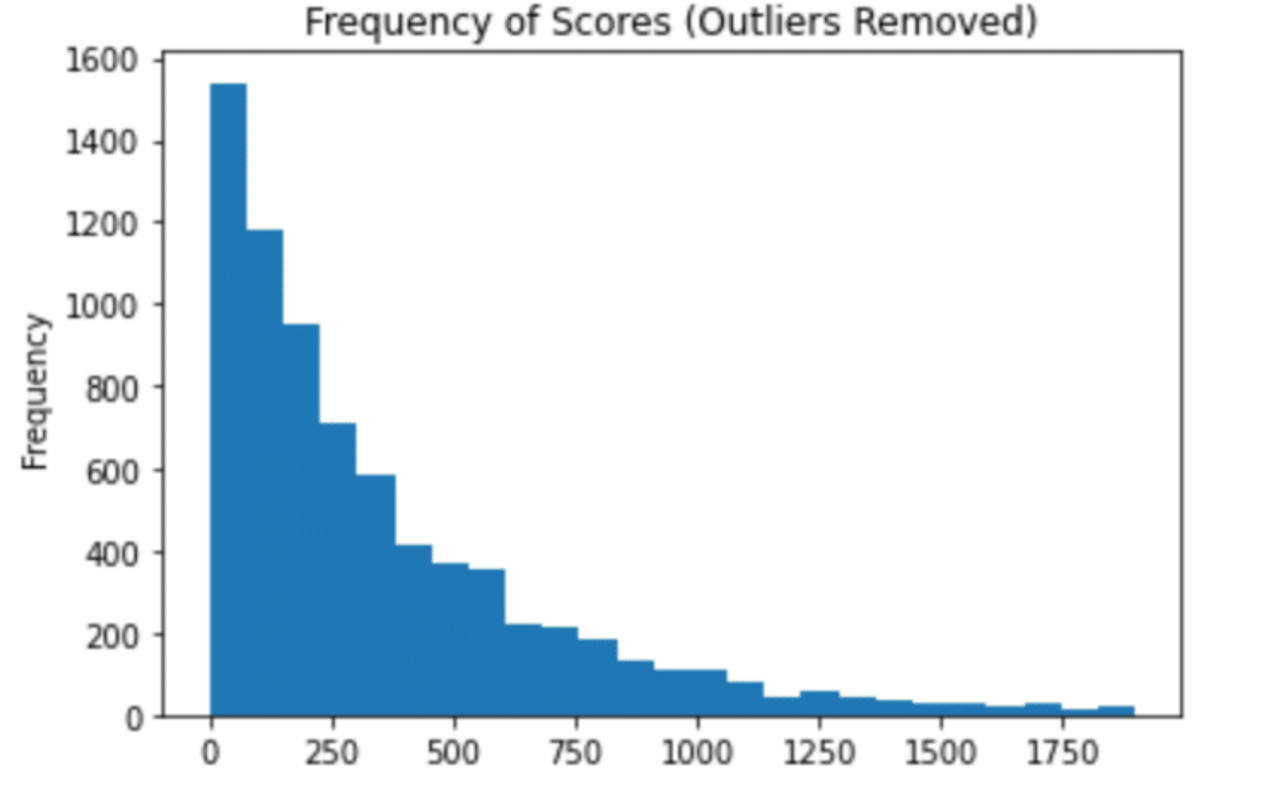}
    \caption{The left plot is the frequency of scores in Georgia dataset. The right plot is the frequency of scores in New Mexico dataset.}
    \label{GAstat}
\end{figure}

The following provides a brief description of some of these variables:
\begin{itemize}
\item Location: Denotes the location of the center of a $600m^2$ grid block where the users measured the connection speed.

\item Download speed: Represents the measured rate of data transfer from a server to the user's device. This is a critical metric as it affects the speed of web page loading and file downloading, and hence is quite noticeable to the user. Download speeds are reported as an average in the grid location. 

\item Upload speed: Corresponds to the measured data transfer rate from the user's device to the server. This metric is crucial for tasks like video calls, cloud file uploads, and user-generated live streaming. Upload speeds are reported as an average in the grid location. 

\item Number of tests at each location: This refers to the number of times tests were conducted in a given grid block. More tests tend to indicate a more reliable measurement score. However, in around $50\%$ of locations, only a single test is conducted.

\item Number of devices used for testing: This refers to the number of distinct devices used for testing in a certain grid block, as the performance can vary across different mobile phones and mobile providers.

\item Score: This variable combines both the download and upload speeds to provide a holistic view of the connection
bandwidth performance.
\end{itemize}
We have mobile connection data from the states of Georgia in the United States Southeast and New Mexico in the United States Southwest. The geographic scatter plot of the connection scores for each dataset is shown in Figure~\ref{s1}. There are 28,587 data points in the Georgia dataset. The frequency of scores is shown in Figure~\ref{GAstat}. There are 7,579 data points in the New Mexico dataset and the frequency of scores is shown in Figure~\ref{GAstat}. The dataset does not include the providers of the devices, which means that this user-collected dataset is a mixture of different internet providers.

Meanwhile, we have removed approximately $2.5\%$ of outliers from the datasets. These outliers are identified as points that deviate significantly from the mean of their $50$-nearest neighbors. Specifically, a point is considered an outlier if the difference between its score and the neighborhood mean exceeds three times the standard deviation of the scores within that neighborhood. This method of identifying outliers is a common practice in data analysis.

\section{Methodology}
Our approach consists of two distinct phases. In the initial phase, we construct a map estimating the score using the kernel regression method. In the subsequent phase, we introduce a novel ensembled spatial conformal prediction method that generates an uncertainty map throughout the region.

\subsection{Self-tuning Bandwidth Kernel Regression}
\label{kr}
Kernel regression is a non-parametric technique to estimate the conditional expectation of a random variable. The objective is to find a non-linear relationship between the input variable and the corresponding output. Kernel regression employs kernel functions, which can capture more complex patterns than linear regression.

For a given dataset with inputs $\mathbf{x}$ and corresponding outputs $\mathbf{y}$, the kernel regression estimate $\hat{y}$ at a new input point $x$ is given by:
\begin{equation} 
\hat{y}(x) = \frac{\sum_{i=1}^{n} K_h(\|x - x_i\|)y_i}{\sum_{i=1}^{n} K_h(\|x - x_i\|)}.
\label{KR}
\end{equation}

Here, $K_h(u) = \frac{1}{h}K(\frac{u}{h})$ is a kernel function, and $h$ is a bandwidth parameter. The kernel function $K(\cdot)$ is often chosen to be a Gaussian kernel, though other choices are also possible. The bandwidth parameter $h$ controls the width of the kernel, and hence the smoothness of the estimated function.

In the Gaussian case, the kernel function $K(\cdot)$ is defined as:
\begin{equation}
K(u) = \frac{1}{\sqrt{2\pi}}\exp\left(-\frac{1}{2}u^2\right).
\end{equation}

Kernel regression estimates the conditional mean function without imposing a parametric form for the functional relationship between predictors and the outcome variable. It allows for flexible, data-driven model specification.

The bandwidth $h$ is a crucial parameter in kernel regression. If $h$ is too small, the estimate will be very rough, capturing too much noise in the data (overfitting). If $h$ is too large, the estimate will be too smooth, not capturing important patterns in the data (underfitting). The choice of an appropriate $h$ often involves cross-validation or some other form of bandwidth selection strategy.

Kernel regression, by its standard definition, utilizes a constant bandwidth $h$ that is applied uniformly across all data points. However, this one-size-fits-all approach might not be the most suitable in scenarios where the data exhibits spatial imbalances, as is the case in our study.

In regions where data points are sparse, the lack of neighboring points could potentially lead to unrepresentative averages and, consequently, inaccurate predictions. To overcome this, we propose the use of a larger bandwidth in such areas, thereby encompassing more points for computation and increasing the representativeness of the estimates.

To this end, we propose the self-tuning bandwidth in the kernel regression framework (STBKR). The self-tuning bandwidth mechanism adapts the bandwidth for each point based on its surrounding density of points. The formula for our self-tuning bandwidth kernel regression can be expressed as follows:
\begin{equation}
\hat{y}(x) = \frac{\sum_{i=1}^{n} K_{h(x)}(\|x - x_i\|)y_i}{\sum_{i=1}^{n} K_{h(x)}(\|x - x_i\|)}.
\end{equation}

In this formula, we let $h(x)=cR_k(x)^2$. Here, $c$ is a parameter that is determined by cross-validation, and $R_k(x)$ denotes the Euclidean distance from a given data point $x$ to its $k$-th nearest neighbor. Notably, in areas where data is sparsely distributed, the distance $R_k(x)$ will be larger, thereby leading to an increased bandwidth $h(x)$.

This strategy of self-tuning bandwidth effectively addresses the issue of spatial imbalance in the dataset. By allowing the bandwidth to adapt based on the local data density, it ensures a more representative sampling of neighbors, leading to more accurate and reliable predictions. Furthermore, the choice of $c$ through cross-validation aids in avoiding overfitting or underfitting, further strengthening the robustness of our regression model.

\subsection{Conformal Prediction}
To begin with, we introduce the basic idea of conformal prediction. Denote the kernel regression model introduced in the previous section as our prediction model $\hat{f}$. The split conformal prediction \citep{vovk2005algorithmic} uses a hold-out set to learn the distribution of the non-conformity score. Assume the hold-out set consists of data $(X_1,Y_1),\cdots,(X_n,Y_n)$, the residuals of the prediction model $\hat{f}$ would be $\varepsilon_i = |Y_i-\hat{f}(X_i)|$ $(i=1,\cdots,n)$.

The prediction interval with confidence $1-\alpha$ at the new point $X_{\text{test}}$ would be $\hat{\mathcal{C}}_{n}^{\alpha}(X_{\text{test}}) = \hat{f}(X_{\text{test}})\pm(\text { the }\lceil(1-\alpha)(n+1)\rceil \text {-th smallest of } \varepsilon_1, \ldots, \varepsilon_n, \infty)$.

In other word, it is saying that the residual $\varepsilon_{\text{test}}$ at the new point $X_{\text{test}}$ should fall in the interval $[0, \text { the }\lceil(1-\alpha)(n+1)\rceil \text {-th smallest of } R_1, \ldots, R_n, \infty]$.

The method guarantees a distribution-free predictive coverage at the target level $1-\alpha$, which means that when the data are independently and identically distributed (i.i.d.), the marginal coverage of prediction region in split conformal prediction satisfies $\mathbb{P}(Y_{\text{test}}\in\hat{\mathcal{C}}_{n}^{\alpha}(X_{\text{test}}))\ge 1-\alpha$. In other words, the prediction interval contains the target point with probability at least $1-\alpha$. The result holds for any sample size $n$, which is called the finite-sample guarantee.

\subsection{Ensemble Spatial Conformal Prediction}
The classic setting of conformal prediction is the i.i.d. or exchangeable setting, which fails to work well when the data has correlations. Moreover, few works consider spatial data that is common in practice. Here we propose a novel ensembled conformal prediction method that is designed for a spatial setting.

\begin{algorithm}[t]
\caption{Ensemble Spatial Conformal Prediction (ESCP)}
\label{alg}
\DontPrintSemicolon  % Uncomment if you don't want semicolons to print at the end of each line

\KwIn{Training data $\left\{\left(x_i, y_i\right)\right\}_{i=1}^N$, prediction algorithm $\hat{f}$, significance level $\alpha$, number of bootstrap models $B$, neighborhood size $K$, batch size $s$, and test data $\left\{\left(x_j, y_j\right)\right\}_{j=1}^{N_t}$}
\KwOut{Ensemble prediction intervals $\{C^{\alpha}\left(x_j\right)\}$ for $x_j$ in test set}

\ForEach{$x_j$ in test set}{
    Initialize $\hat{\bm{\varepsilon}}=\{\}$ as an ordered set\;
    Select $K$-nearest neighbors of $x_j$ in training data as a set $N(x_j)=\{x_{n_1},\cdots,x_{n_K}\}$\;
    \For{$b = 1$ \KwTo $B$}{
        Sample with replacement an index set $D_b(x_j)=(n_{i_1},\cdots,n_{i_s})$ from $N(x_j)$\;
        Compute $\hat{f}^b=\hat{f}_{D_b(x_j)}$\;
    }
    \For{$i = 1$ \KwTo $K$}{
        Compute $\hat{f}_{-i}(x_{n_i})=\text{mean}(\hat{f}^b(x_{n_i}), n_i \notin D_b)$\;
        Compute $\hat{\varepsilon}_{-i} = |\hat{f}_{-i}(x_{n_i})-y_{n_i}|$\;
        $\hat{\bm{\varepsilon}}=\hat{\bm{\varepsilon}} \cup \{\hat{\varepsilon}_{-i}\}$\;
    }
    Fit quantile regression $\hat{Q}$ on $\hat{\bm{\varepsilon}}$\;
    Compute $\omega^{\alpha}(x_j)=(1-\alpha)$ quantile of $\hat{\bm{\varepsilon}}$\;
    Return $C^{\alpha}\left(x_j\right)=[\hat{f}_{D_0(x_j)}(x_j)-\omega^{\alpha}(x_j), \hat{f}_{D_0(x_j)}(x_j)+\omega^{\alpha}(x_j)]$\;
}
\end{algorithm}
In Algorithm \ref{alg}, the prediction model $\hat{f}$ we use is the self-tuning bandwidth kernel regression method in Section \ref{kr}. We enhance this method by using an ensemble approach, which significantly boosts both accuracy and robustness. The enhancement is also effective in handling spatial data. To this end, for each data point, we identify its $K$-nearest neighbors to define a local area, which reflects the spatial dependencies relevant to that point. 

For training the model, we employ the bootstrap technique, randomly sampling several points with replacement from this localized dataset. These sampled points serve as the training data for fitting the predictive model. This process is not just a one-off but is iterated multiple times, yielding a diverse collection of predictors. Each iteration contributes unique perspectives to the ensemble, enhancing the model's overall predictive power and stability. Based on these predictors, we use an ``leave-one-out'' estimator $\hat{f}_{-i}$ that is related to the Jackknife+ procedure \cite{barber2021predictive} to predict the signal quality. The procedure avoids splitting the dataset to fit the prediction model, maintaining the statistical efficiency while still being computationally efficient. 

Once we compute the non-conformity scores (essentially, the residuals) for each point within the local neighborhood, we proceed to fit a quantile regression model $\hat{Q}$. Recent studies \cite{romano2019conformalized, xu2023sequential} have found that replacing empirical quantile with quantile regression can capture the dependency between residuals and often leads to tighter confidence regions. There is no restriction on the quantile regression method used in Algorithm \ref{alg}. In general, quantile regression algorithms rely on minimizing the pinball loss for specific regression algorithms. However, we typically want the method to be computationally efficient and require few tunings of hyperparameters. In the experiment, we chose the Quantile Random Forest (QRF) \cite{meinshausen2006quantile} that is fast and easy to use.

In quantile regression, the model learns the conditional distribution of $Y|X$, where $Y$ is the response and $X$ is the feature. In our setting, we train the QRF on the residuals in the neighborhood to leverage the dependency. To be precise, for a data point located at $x$, we find its $K$-nearest neighbor set $N(x)=\{x_{n_1},\cdots,x_{n_K}\}$, where the set is sorted from geographically closest to furthest. Based on the prediction model $\hat{f}$, we are able to compute the non-conformity scores $\varepsilon_{n_i}$, and we denote the score at point $x$ as $\varepsilon_{n_0}$. Then we let
\begin{equation}
X=[\varepsilon_{n_1},\cdots,\varepsilon_{n_K}], Y=\varepsilon_{n_0}.
\end{equation}
For each point in the training set, we are able to construct such a neighborhood and compute corresponding non-conformity scores. Then we can train QRF a large neighborhood of $x$ in the training set. When re-fitting QRF at each prediction point, we are capturing the local spatial dependency between residuals.

\section{Numerical Experiment}
We initiate our numerical experiment by partitioning the data into training and test sets with a split of $80\%$ and $20\%$ respectively. To optimize the parameters $k$ and $c$ in the self-tuning bandwith kernel regression, we implement a 5-fold cross-validation on the training set. We select the combination that minimizes the $\ell_2$ prediction error. The choice is $k=10$ and $c=0.01$.

For the ensemble spatial conformal prediction, we select the bootstrap number $B=50$, and the batch size $s=100$. A larger choice of $B$ and $s$ will lead to high accuracy and robustness. In our experiment, the current choice is able to reach promising results while reducing computational costs to a large extent. We set $\alpha$ to be $0.2$, which means that the intervals should have at least $1-\alpha=80\%$ probability of coverage. This significance level produces a tighter region, whereas a higher significance level ensures greater coverage, though it may lead to a conservatively larger prediction interval.

Although neither Georgia nor New Mexico is rectangular in shape, we have included additional data to extend their areas to a rectangular format for the ease of displaying results.

\begin{figure}[t]
    \centering
    \includegraphics[scale=0.38]{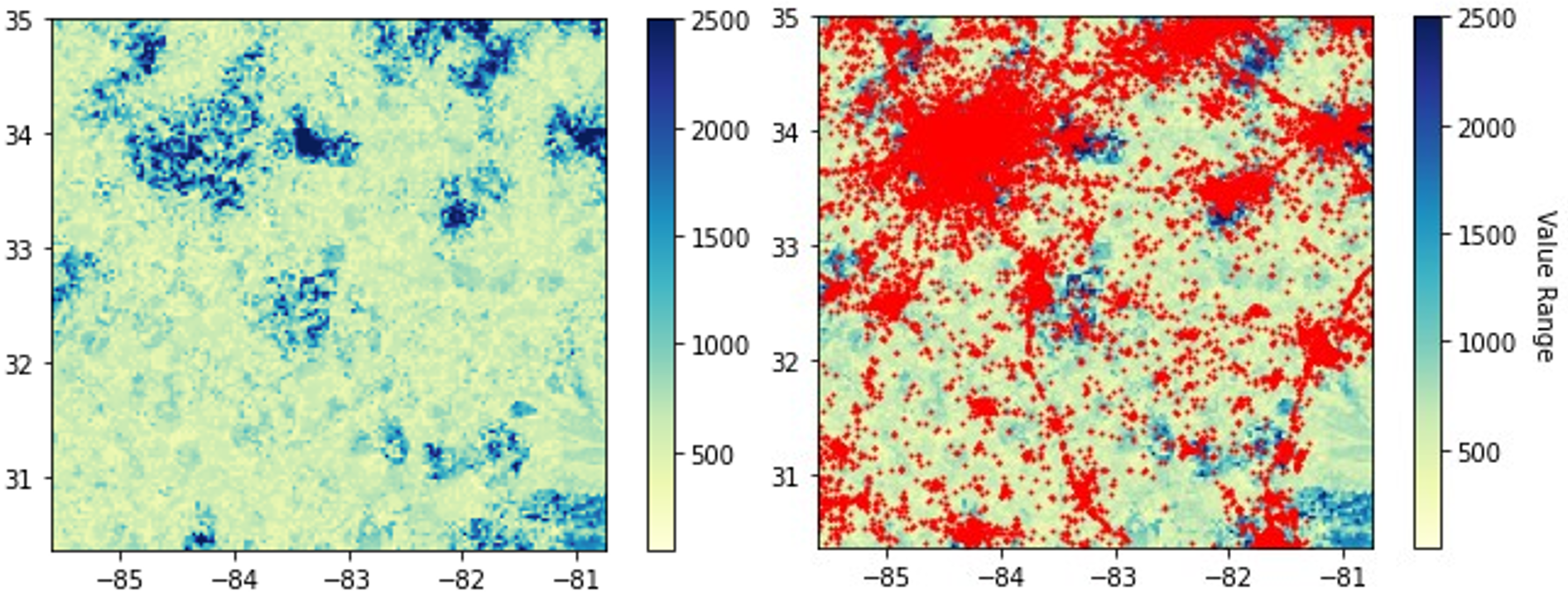}
    \caption{The left plot is the uncertainty map of Georgia, and the magnitude shows the width of the prediction interval of score in the dataset. The significance level $1-\alpha=0.8$. The plot on the right shows the locations of training data.}
    \label{uq1}
\end{figure}

\begin{figure}[t]
    \centering
    \includegraphics[scale=0.38]{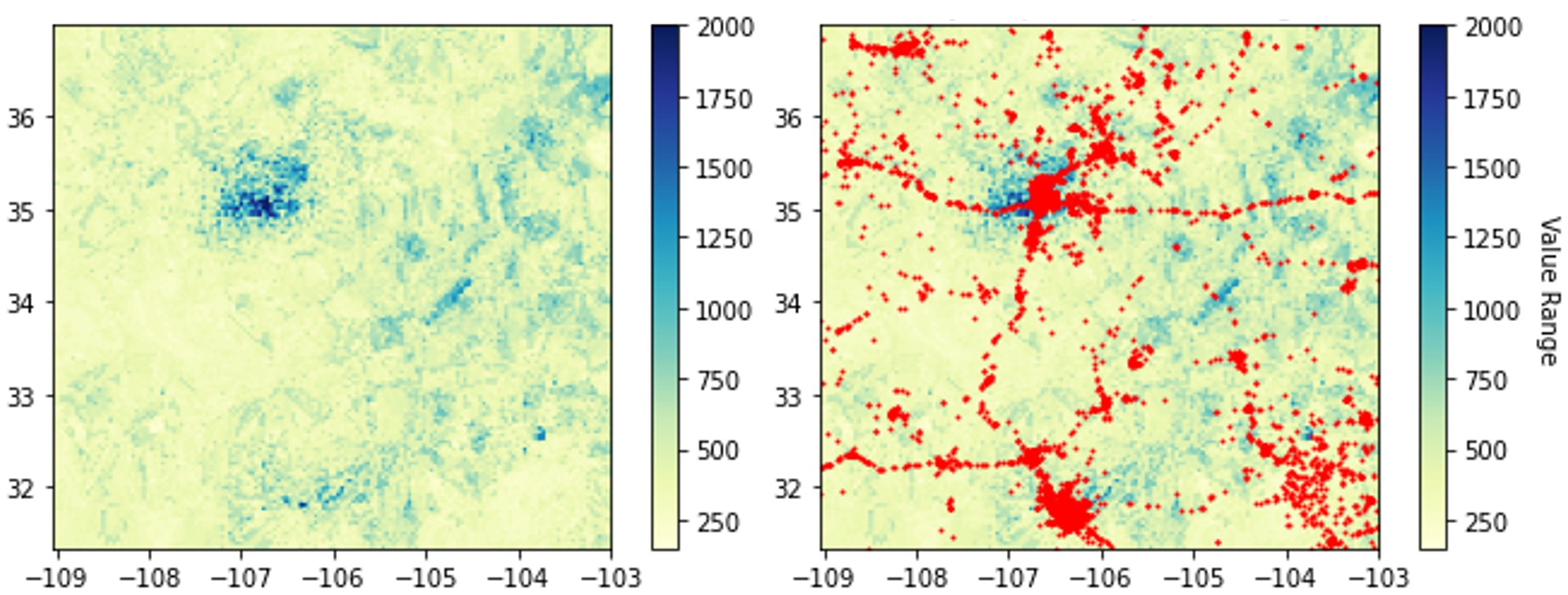}
    \caption{Uncertainty map of New Mexico, and the magnitude shows the width of the prediction interval of score in the dataset. The significance level $1-\alpha=0.8$. The plot on the right shows the locations of training data.}
    \label{uq2}
\end{figure}

In the experiment, we use our ensemble spatial conformal prediction method on the Georgia dataset and New Mexico dataset. We compare the performance of our method with two baseline methods. The first one is the standard split conformal prediction. The second one is EnbPI \cite{xu2021conformal}, which is also a conformal prediction method that utilizes ensemble predictors. We split the dataset into $80\%-20\%$

We compare the coverage rate on the test set and average width of prediction intervals on the test data. The results are presented in Table \ref{GAT} and \ref{NMT}. From the table we can see that all the three methods achieve valid coverage ($0.8$) on the test data. Our proposed ESCP method generates a tighter prediction region compared to the other two methods.

From Figure \ref{uq1} and \ref{uq2}, we observe that our method identifies several small areas exhibiting high uncertainty. High uncertainty indicates that the prediction model lacks confidence in its estimates, meaning the actual values could significantly deviate from the predictions. It is apparent that uncertainty peaks in urban regions where extensive measurements are recorded. This heightened uncertainty in densely populated areas arises from factors such as network congestion, signal obstructions from buildings, and interference from other electronic signals. Additional data collection may not significantly enhance estimation accuracy in these areas due to the inherently high variability of internet speeds. 

On the other hand, certain rural areas also display elevated uncertainty due to limited data availability. In contrast, rural regions typically experience less interference and network congestion, contributing to more stable internet speeds. For instance, in the New Mexico dataset shown in Figure \ref{s1}, numerous data points are collected along highways in rural areas. Interestingly, our uncertainty quantification map indicates low uncertainty along these highways, suggesting more stable internet conditions there. However, other rural regions with scant data collection do show higher uncertainty. This observation underscores the need for targeted data collection in these rural areas, as acquiring more data could significantly refine our internet speed estimations, given the inherently lower variability in these regions.

\begin{table}[t]
\caption{Coverage and width of prediction interval on Georgia dataset.}
\label{GAT}
\vskip 0.15in
\begin{center}
\begin{small}
\begin{sc}
\begin{tabular}{lcccr}
\toprule
   & split CP & EnbPI & ESCP  \\
\midrule
Coverage   & 83.4 & 82.4 & 80.8\\
Width & 1501 & 1677 & 1180\\
\bottomrule
\end{tabular}
\end{sc}
\end{small}
\end{center}
\end{table}

\begin{table}[t]
\caption{Coverage and width of prediction interval on New Mexico dataset.}
\label{NMT}
\vskip 0.15in
\begin{center}
\begin{small}
\begin{sc}
\begin{tabular}{lcccr}
\toprule
   & split CP & EnbPI & ESCP  \\
\midrule
Coverage   & 84.6 & 82.9 & 82.6\\
Width & 780 & 839 & 701\\
\bottomrule
\end{tabular}
\end{sc}
\end{small}
\end{center}
\end{table}

\section{Discussion}
This paper explores the challenge of quantifying uncertainty in cellular network speeds across diverse geographic regions. We have demonstrated that the variability of internet speeds significantly complicates accurate speed prediction due to factors such as high user density, structural barriers, and interference from other signals. We propose a novel Ensemble Spatial Conformal Prediction (ESCP) method that not only complements the prediction but also enables more informed decisions about where to focus data collection efforts.

We must emphasize that the conformal prediction method we utilize is based on the self-tuning bandwidth kernel regression model. The precision of this model greatly influences the prediction interval's width, or its tightness. Accurately estimating mobile internet speeds is particularly challenging due to their high variability. There is limited research focused on creating comprehensive coverage maps across entire states, and traditional models often struggle with noisy and incomplete datasets.

Using the Ookla dataset, we have generated uncertainty maps for Georgia and New Mexico. These maps reveal greater uncertainty in densely populated areas and in rural regions lacking sufficient data. This outcome confirms our method's capability to identify areas of uncertainty. Furthermore, our tests on both datasets demonstrate that our approach achieves valid coverage and produces tighter prediction intervals than those generated by baseline methods.

By identifying areas where the prediction model is uncertain, we can guide researchers in manually collecting high-quality data. Gathering more data in underrepresented areas will enhance both the accuracy of our estimations and the tightness of our prediction intervals. Given that our method has empirically shown to provide valid coverage, actively collecting data in these areas will undoubtedly refine the prediction intervals and guarantee accuracy. This active sampling strategy significantly boosts the quality of the data collected.

\begin{acks}
This work was partially supported by National Science Foundation (NSF) CNS-2220387.
\end{acks}

\appendix

\section{Ethics}
Cellular measurement data has the potential to reveal an individual’s location or movement over time. The data used in our work helps protect against this with spatial aggregation into $600m^2$ regions and time aggregation over multiple years of data collection.
% Note from the CFP that this section must include a statement about
% ethical issues; papers that do not include such a statement may be
% rejected.

%%%%%%%%%%%%%%%%%%%%%%%%%%%%%%%%%%%%%%%%%%%%%%%%%%%%%%%%%%%%%%%%%%%%%%%%%%%%
% We're in the endgame now

\bibliographystyle{ACM-Reference-Format}
\bibliography{refs}

\end{document}